\documentclass[aps,pre,reprint,groupedaddress]{revtex4-1}

\usepackage{amsmath}
\usepackage{amssymb}
\usepackage[figuresright]{rotating}
\usepackage{color}
\usepackage{graphicx}
\usepackage{subfigure}
\usepackage[sort&compress]{natbib}
\newcommand{\abs}[1]{\left| #1 \right|} 
\newcommand{\avg}[1]{\left< #1 \right>} 
 
\newcommand{\ket}[1]{\left| #1 \right>} 
\newcommand{\bra}[1]{\left< #1 \right|} 
\newcommand{\braket}[2]{\left< #1 \vphantom{#2} \right|
 \left. #2 \vphantom{#1} \right>} 

\begin{document}


\title{Quantum field theory treatment of magnetic effects on the spin and orbital angular momentum of a free electron} 

\author{P. Kurian}
\email[Corresponding author. Electronic mail: ]{pkurian@gmx.com.}
\affiliation{National Human Genome Center, Howard University College of Medicine, Washington, DC, USA}

\author{C. Verzegnassi}
\affiliation{Department of Chemistry and Environmental Physics, University of Udine, Udine, Italy}

\date{\today}

\begin{abstract}
We consider in a quantum field theory framework the effects of a classical magnetic field on the spin and orbital angular momentum (OAM) of a free electron. We derive formulae for the changes in the spin and OAM due to the introduction of a general classical background field. We consider then a constant magnetic field, in which case the relevant expressions of the effects become much simpler and conversions between spin and OAM become readily apparent. An estimate of the expectation values for a realistic electron state is also given. Our findings may be of interest to researchers in spintronics and the field of quantum biology, where electron spin has been implicated on macroscopic time and energy scales.
\end{abstract}


\maketitle 


Conversion of electron spin into orbital angular momentum (OAM) and its harnessing for spintronic devices has been stressed in the recent literature \cite{Karimi1, Chekhovich, Karimi2, Karimi3, Karimi4, Spinhall}. One essential aspect of the different processes is the spin orientation of the constituent electrons, which can be considered as free. In particular, variations in the electron spin polarization can be discriminated by interactions with atomic states of strongly coupled spin and OAM, which produces a change in OAM of the electrons \cite{Spinhall}. 

This reciprocal relationship between electron spin and OAM is made very clear in the special case of electron vortex beams \cite{Karimi1, Karimi2, Karimi3, Karimi4}. Researchers have developed a device that induces spin flips by a suitable magnetic field and converts the corresponding spin angular momentum variation into OAM of the electron beam itself by exploiting a geometric Berry phase arising from the spin manipulation. Such technology can be applied to a spin-polarized electron beam to generate a so-called vortex beam carrying OAM, or to an unpolarized input to act as a spin-polarization filter.

Quite separately, magnetic field effects on spin polarization in living systems have long been studied \cite{Presman}. The recent surge of interest in the quantum coherent mechanisms of the avian compass \cite{Gauger, Wiltschko, Cai, JNBandyopadhyay} indicates the primacy of electron spin in orienting migratory birds to the weak magnetic fields of the Earth. Just last year, it was demonstrated that general anesthetics rapidly increase the electron spin content in \textit{Drosophila} fruit flies \cite{Turin}, suggesting that these molecules perturb electron currents in cells and that their efficacy may be affected by magnetic fields. Thus, the study of such field effects on spin has implications both in the manufacture of new devices and in fundamental physical explanations of biological processes.

We have described in this letter the effects that a magnetic field would have on a free electron's spin and OAM in the framework of quantum field theory (QFT).  Our choice is motivated by the fact that in QFT both spin and OAM are precisely defined at the outset and only depend on the quantum fermion field, whereas in quantum mechanics spin arises as an ad-hoc addition to the OAM. The definition of these quantities are a consequence of the choice of the Lagrangian and the imposition of rotational invariance in the application of Noether's theorem. Computing the change in spin and OAM due to a magnetic field in this framework will show a systematic and unavoidable correlation, since both changes will be fixed by the variation of the same field. We shall clarify this statement in what follows.

Our starting point will be the theoretical expressions for the spin $\vec{S}$ and OAM $\vec{L}$ in QFT for a free electron system described by the usual four-component fermion field $\psi(\mathbf{x})$, with $\mathbf{x} \equiv (x_0, \vec{x})= (x_0, x_1, x_2, x_3)$:
\begin{eqnarray}
\vec{S} =& (S_{23}, S_{31}, S_{12}) = (S_1, S_2, S_3), \\ \label{spin}
S_{ab} =& \int d\mathbf{x}\,\, \psi^\dagger(\mathbf{x}) \,\, \frac{1}{2} \sigma_{ab} \,\, \psi(\mathbf{x}), \\
\sigma_{ab} =& i \gamma^a \gamma^b, \\ \label{OAM}
\vec{L} =& \int d\mathbf{x}\,\, \psi^\dagger(\mathbf{x}) \,\, \vec{\ell} \,\, \psi(\mathbf{x}), \\ \label{ell}
\vec{\ell} =&  (-i) (\vec{x} \times \vec{\nabla}).
\end{eqnarray}

We now introduce a classical electromagnetic field $\mathbf{A}$, described in the usual way by the four potentials $A_0, A_1, A_2, A_3$, and follow the Dirac prescription that uses the replacement of the ordinary derivative by the so-called covariant derivative:
\begin{equation}
\partial_\mu \rightarrow \partial_\mu - ieA_\mu,
\end{equation} 
where $e$ is the fermion (electron) charge. We want to examine the effects when we introduce the classical electromagnetic field at a fixed time $x_0=0$.

Requiring that the electron field must satisfy the Dirac equation leads to changes in the four components $\psi_1, \psi_2, \psi_3, \psi_4$ of the spinor field. These changes have already been derived \cite{Verzegnassi} and denoted by the expressions $\Delta_{\mathbf{A}} \psi_i$: 
\begin{eqnarray} \nonumber
\Delta_{\mathbf{A}}\psi_1 &= \frac{\abs{e}}{m} \left(A_0\psi_3+A_1\psi_4-iA_2\psi_4+A_3\psi_3\right),\\ \nonumber
\Delta_{\mathbf{A}}\psi_2 &= \frac{\abs{e}}{m} \left(A_0\psi_4+A_1\psi_3+iA_2\psi_3-A_3\psi_4\right),\\ \nonumber
\Delta_{\mathbf{A}}\psi_3 &= \frac{\abs{e}}{m} \left(A_0\psi_1-A_1\psi_2+iA_2\psi_2-A_3\psi_1\right),\\  \label{ClaudioDeltas}
\Delta_{\mathbf{A}}\psi_4 &= \frac{\abs{e}}{m} \left(A_0\psi_2-A_1\psi_1-iA_2\psi_1+A_3\psi_2\right).
\end{eqnarray}

We choose to concentrate our study on the changes in $\vec{S}$ and $\vec{L}$ due to the purely magnetic potentials $\vec{A}$. Denoting these effects on a given quantity by $\Delta_{\vec{A}}$, we have derived them by replacing the four components $\psi_i$ with their modified expressions
\begin{equation}
\psi_i \rightarrow \psi_i + \Delta_{\vec{A}}\psi_i
\end{equation}
and inserting them in Equations (\ref{spin}) and (\ref{OAM}). In the calculations we have only retained the first-order corrections to the quantities, which corresponds to retaining the lowest terms in the perturbative electron charge expansion.

The modifications of the spin components are given by:
\begin{equation}
\Delta_{\vec{A}} \vec{S} = \abs{e} \int d\vec{x} \,\, \left[\vec{A} \times \vec{\rho}_E \right ],
\end{equation}
where $\vec{A}$ is the magnetic potential and $\vec{\rho}_E$ is defined as 
\begin{equation}
\vec{\rho}_E = \frac{i}{m_e}\psi^\dagger \vec{\gamma} \psi
\end{equation}
and identified as a kind of electric dipole because the change in energy density of a free electron produced by an electric field $\vec{E}$ has the form $\vec{\rho}_E \cdot \vec{E}$ \cite{Verzegnassi}. The shift in the OAM has a more complicated expression, which reduces to something more reasonable when the calculations are performed for the various components of $\vec{L}$. Here we give the formula that provides the effect on the third component of the OAM:
\begin{equation}
\Delta_{\vec{A}}L_3 = -i\abs{e}\int d\vec{x} \left[\rho_{E1} (\ell_3A_1) + \rho_{E2} (\ell_3 A_2) + \rho_{E3} (\ell_3 A_3) \right],
\end{equation}
where $\ell_1, \ell_2, \ell_3$ are defined by Equation (\ref{ell}). The other components can be derived analogously. As one can see, the overall effect on $\vec{L}$ is still completely fixed by the values of $\vec{A}$ and $\vec{\rho}_E$, in perfect analogy with the effect on $\vec{S}$. 

We now want to compute the magnetic effect on the third component of the spin:
\begin{equation} \label{DeltaS}
\Delta_{\vec{A}}S_3 = \abs{e}\int d\vec{x} \,\, \left[A_1\rho_{E2} - A_2 \rho_{E1} \right].
\end{equation}
To derive more applicable conclusions, we now consider the realistic case of a constant magnetic field $\vec{\mathcal{H}}$, which we choose to be oriented along the $z$ axis, so that $\vec{\mathcal{H}} = (0,0, \mathcal{H}_3)$. The expressions of the magnetic potentials are then fixed by 
\begin{equation}
\vec{A}(\vec{x})= \frac{1}{2} \left(\vec{\mathcal{H}} \times \vec{x} \right),
\end{equation}
which leads to the following expression:
\begin{equation}
\Delta_{\vec{A}(\mathcal{H}_3)}S_3 = -\abs{e} \frac{\mathcal{H}_3}{2} \int d\vec{x} \,\, \left[x_2 \rho_{E2}  + x_1 \rho_{E1} \right].
\end{equation}

One can similarly perform the calculation of $\Delta_{\vec{A}(\mathcal{H}_3)}L_3$ and find the result
\begin{equation} \label{consangmom}
\Delta_{\vec{A}(\mathcal{H}_3)}L_3 = -\Delta_{\vec{A}(\mathcal{H}_3)}S_3.
\end{equation}
This is in fact an obvious consequence of the conservation of total angular momentum. In the chosen example, there is a symmetry under rotations in a plane perpendicular to the $z$ axis. As a consequence, the third component of the total angular momentum $\vec{J} = \vec{L} + \vec{S}$ must remain conserved, which means $\Delta_{\vec{A}(\mathcal{H}_3)}J_3=0$ and thus $\Delta_{\vec{A}(\mathcal{H}_3)}L_3 + \Delta_{\vec{A}(\mathcal{H}_3)}S_3=0$ as stated above in Equation (\ref{consangmom}). 

The conservation of total angular momentum in this special case can be derived from the general expression for a constant magnetic field in any direction. A straightforward calculation leads to the following result:
\begin{equation} \label{DeltaJ}
\Delta_{\vec{A}}J_i= - \frac{\abs{e}}{2} \int d\vec{x} \,\, \left[\varepsilon_{ijk} \, \mathcal{H}_j \left(\vec{x} \times \vec{\rho}_E \right)_k \right],
\end{equation}
where $\varepsilon_{ijk}$ is the Levi-Civita tensor. One possible conclusion that could be derived from Equation (\ref{DeltaJ}) is that QFT predicts possible conversions of spin into OAM. More importantly, one can derive in this framework the values of these changes. 

We have computed the expectation values of the changes in spin and OAM for a realistic electron state. More precisely, we have considered a free electron state of momentum $\vec{k}=(0,0,k_3)$ along the $z$ axis, in a linear superposition of spin eigenstates:
\begin{equation}
\ket{\Psi(\vec{k})} = \lambda_{+} \ket{\uparrow, \vec{k}} + \lambda_{-} \ket{\downarrow, \vec{k}}.
\end{equation} 
We choose the following completeness relation to hold: $\abs{\lambda_+}^2 + \abs{\lambda_-}^2 = 1$. In the conventional treatment, 
\begin{equation}
 \ket{s, \vec{k}} = \sqrt{2E_0} a^\dagger_{s, \vec{k}} \ket{0},
 \end{equation}
where $s=\uparrow,\downarrow$, $\ket{0}$ is the vacuum state, $E_0 = \sqrt{|\vec{k}|^2+m_e^2}$, and $a^\dagger_{s, \vec{k}}$ is the fermionic creation operator for a state of spin $s$ and momentum $\vec{k}$. This creation operator is the one that appears in the standard Fourier transform of $\psi^\dagger (\mathbf{x})$ \cite{Peskin}.
 
We have calculated the expectation value of $\Delta_{\vec{A}}S_3$ in the state $\ket{\Psi(\vec{k})}$, conventionally defined as 
\begin{equation}
\avg{\Delta_{\vec{A}}S_3} = \frac{\bra{\Psi(\vec{k})}\Delta_{\vec{A}}S_3\ket{\Psi(\vec{k})}}{\braket{\Psi(\vec{k})}{\Psi(\vec{k})}}.
\end{equation}
The numerator can be calculated starting from the expression derived in Equation (\ref{DeltaS}). 

An important feature of the calculation is the size of the integration volume. It seems reasonable that the integration be performed in the region of space where $A_1, A_2 \neq 0$, which is fixed by the scale $d$ of the experimental apparatus used to generate the magnetic field. We assume that $\avg{A_1}, \avg{A_2}$ remain reasonably close to their average values in this volume so that they can be extracted from the integral as numbers. This allows us to compute the expectation value of $\Delta_{\vec{A}}S_3$:
\begin{equation}
\avg{\Delta_{\vec{A}}S_3}=-2\frac{\abs{e}}{m_e} \frac{|\vec{k}|}{E_0}\left[\avg{A_1}\operatorname{Re}(\lambda_+ \lambda_-^*) - \avg{A_2}\operatorname{Im}(\lambda_+ \lambda_-^*) \right].
\end{equation}
For our constant magnetic field along the $z$ axis, we would obtain 
\begin{equation}
\avg{A_1} \approx -\frac{1}{2}\mathcal{H}_3 d, \quad \avg{A_2} \approx \frac{1}{2}\mathcal{H}_3 d.
\end{equation}
To obtain a numerical estimate, we use characteristic values for the magnetic field of $O(10^{-5})$ Tesla and for the experimental apparatus $d=1$ meter. Thus, in natural Planck units and assuming $|\vec{k}| \gg m_e$ so that $|\vec{k}|/E_0 \approx 1$, we obtain
\begin{equation} \label{S3est}
\avg{\Delta_{\vec{A}}S_3} \approx \operatorname{sgn}(\mathcal{H}_3)(0.029)\left[\operatorname{Re}(\lambda_+ \lambda_-^*)+\operatorname{Im}(\lambda_+ \lambda_-^*)\right],
\end{equation}
which suggests non-negligible changes in the spin, on the order of a few percent. 

We now present a few comments on Equation (\ref{S3est}). Numerically, $\abs{\operatorname{Re}(\lambda_+ \lambda_-^*)+\operatorname{Im}(\lambda_+ \lambda_-^*)}$ attains a maximum value equal to $\frac{1}{2}$ for several electron states of initially mixed polarization. In these cases, the value of the electron spin change is primarily determined by the external magnetic field strength and direction. However, if the initial electron is fully polarized, either in the right-handed state $\abs{\lambda_+}=1$ or in the left-handed state $\abs{\lambda_-}=1$, the magnetic field effect will be vanishing. The magnetic field effects on the first and second components of the spin $\vec{S}$ will also be vanishing, because the rotational invariance about the $z$ axis demands that the spin-OAM conversion (\ref{consangmom}) comes from the $S_3$ component alone. 

We have shown that in a QFT framework the interaction with an external classical magnetic field will generate changes in the spin and OAM of a free electron, and that these changes are correlated and can be simply calculated. These conclusions might be of some interest to biological researchers, particularly in quantum biology, as a fractal-like hierarchy may exist between the spin polarization of elementary matter and the physiological function of complex living organisms \cite{Vitiello, Mandelbrot, Bohm}. The possibility that the holographic nature of reality extends to biology is in our opinion highly relevant, and we look forward to collaborating with interested experimental groups. \\

\begin{acknowledgments}
The authors would like to thank G. Dunston of the National Human Genome Center for her support and encouragement of this work, and E. Spallucci for several useful discussions.
\end{acknowledgments}

\end{document}